\documentclass[graybox]{svmult}

\usepackage{type1cm}
\usepackage{graphicx}
\usepackage{newtxtext}
\usepackage[varvw]{newtxmath}

\graphicspath{{Figures/}{Images/}}
\usepackage{adjustbox}
\usepackage{url}
\usepackage{hyperref}

\usepackage[ruled,vlined,linesnumbered,resetcount]{algorithm2e}
\usepackage{listings}
\usepackage{tikz}
\tikzset{num/.style={anchor = center, inner sep = 2, font = \small}}
\usepackage{pgfplots}
\usepackage{comment}
\usepackage{xcolor}

\definecolor{HighlightColor}{rgb}{0.462745098, 0.725490196, 0.000000000}
\definecolor{HighlightColor2}{rgb}{0.945, 0.349, 0.373}
\definecolor{HighlightColor3}{rgb}{0.349, 0.604, 0.827}
\definecolor{HighlightColor4}{rgb}{0.976, 0.651, 0.353}

\definecolor{Emerald}{rgb}{0.31, 0.78, 0.47}
\definecolor{MidnightBlue}{rgb}{0.1, 0.1, 0.44}
\definecolor{YellowOrange}{rgb}{0.95, 0.52, 0.0}
\definecolor{Magenta}{rgb}{1.0, 0.0, 1.0}
\definecolor{Salmon}{rgb}{1.0, 0.57, 0.64}
\definecolor{LimeGreen}{rgb}{0.462745098, 0.725490196, 0.000000000}

\hypersetup{colorlinks=true,allcolors=HighlightColor}

\begin{document}
\title*{Quasi-Monte Carlo Algorithms (not only) for Graphics Software}
\author{Alexander Keller, Carsten W\"achter, and Nikolaus Binder}
\institute{Alexander Keller \email{akeller@nvidia.com}
\and Carsten W\"achter \email{cwachter@nvidia.com}
\and Nikolaus Binder \email{nbinder@nvidia.com} \at NVIDIA, Fasanenstr. 81, 10623 Berlin, Germany}

\maketitle

\abstract*{-- insert copy --}

\abstract{Quasi-Monte Carlo methods have become the industry standard in computer graphics.
For that purpose, efficient algorithms for low discrepancy sequences are discussed. In
addition, numerical pitfalls encountered in practice are revealed. We then take a
look at massively parallel quasi-Monte
Carlo integro-approximation for image synthesis by light transport simulation. 
Beyond superior uniformity, low discrepancy points
may be optimized with respect to additional criteria, such as
noise characteristics at low sampling rates or the quality
of low-dimensional projections.}


\section{Introduction}

Physically based image synthesis~\cite{PBRT} consists of computing pixel colors as
functionals of the solution of a Fredholm integral equation that models
light transport. The simulations are based on sampling light transport paths
and summing up their contributions. In early computer graphics, Monte Carlo methods
using random samples have been applied. Today,  quasi-Monte Carlo methods
have become the industry standard~\cite{NutshellQMC}
and are an active domain of research~\cite{MyFavoriteSamples}.

There are several reasons for this development. First of all, the points of
low discrepancy sequences as used for sampling in quasi-Monte Carlo methods
are much more uniformly distributed than random numbers ever can be~\cite{Nie:92},
which improves efficiency. In addition, deterministic low discrepancy sequences
allow for reproducible and massively parallel computation without
the risk of inefficiencies. In fact, their extensibility lends itself to
progressive, adaptive, and consistent algorithms.

In the following, efficient implementations of sampling algorithms for quasi-Monte
Carlo methods as used in industry are presented and discussed. For the example
of photorealistic image synthesis, which amounts to numerical integro-approximation
computing millions of integrals in parallel, a new sampling scheme is proposed.
The simple algorithm excels at visual quality, especially at low sampling rates
and hence is a perfect candidate for real-time light transport simulation.

\section{Implementation of Low Discrepancy Sequences}

Digital $(t, s)$-sequences and rank-1 lattice sequences are the two most popular
approaches to generate low discrepancy sequences.
While implementations may seem straightforward, there are several practical issues
that deserve attention.

Today's speed of computing is dominated by the cost of moving data
rather than the cost of arithmetic operations. Hence, replacing memory
accesses by computations may be more efficient than a table lookup. Yet, a smaller
code size and more efficient code may result from the use of the lookup tables.
Balancing these competing goals, both data
and code will remain in the respective processor caches, which may
dramatically improve performance. 

\subsection{Floating Point Conversion} \label{Sec:fp}

Although constructed over finite fields, low discrepancy sequences are points in
the $s$-dimensional unit cube $[0, 1)^s$ whose components are
represented by floating point numbers~\cite{goldberg91}. Note that the set of
floating point numbers in the unit interval $[0, 1)$ is not uniformly distributed.
Due to the representation by a mantissa and an exponent, floating point
numbers are more uniformly distributed the closer they are to zero.
Collisions are another consequence of this representation. For example, mapping
integers to the unit interval by dividing by the maximum value of their range
will ultimately result in cases where different integers become mapped to the same
floating point number.

Another issue is that type conversion and rounding may yield
floating point numbers outside the unit interval. For example,
converting a double precision floating point number close to one in the unit interval to single
precision may result in a one. This in turn may cause errors
in applications that rely on the right-open unit interval $[0, 1)$, where one is excluded,
for example, computing the logarithm $\log(1-x)$ of a component $x$
of a low discrepancy sequence.

To reduce cancellation when adding floating point numbers
of sufficiently different magnitudes, computations should remain in
integers or finite fields as long as possible, including randomization
methods like scrambling and Cranley-Patterson-rotations~\cite{Owen:98}.
For that purpose,
the function \texttt{map\_u32\_to\_unifloat} in Algorithm~\ref{Fig:CodeRadicalInversion} maps a uniformly distributed integer $u \in [0, 2^{32})$ to the best 32-bit floating point representation in the half-open interval [0, 1).
It avoids precision issues of a division-based mapping by explicitly setting bits to appropriate values and is inspired by the concepts found in~\cite{fpPRN,Walker1974FastGO}.
Counting the number of leading zeros determines the smallest power of two $B$ that is greater than the unsigned integer input $u$.
This upper bound defines the exponent $e$ of the floating point number as $B = 2^e$.
For IEEE 754 32-bit floating point numbers with an 8-bit exponent, the bias of the exponent is $2^7 -1 = 127$.
Incorporating this bias, for $z$ leading zeros of the input value the exponent for the target interval becomes $126 - z$.
For architectures on which counting the number of leading zeros of zero is either undefined or does not not result in 32, we introduce a special case and map zero explicitly to zero, which in addition avoids that zero is mapped to the smallest possible floating point number.
We also add a special case for an input value of one, for which the algorithm would shift left by 32 bits. A left shift by the number of bits of the type is undefined behavior according to the C++ standard\footnote{\url{https://en.cppreference.com/w/cpp/language/operator_arithmetic}}.
The leading one of the mantissa of IEEE 754 floating point numbers is implicit.
Therefore, we first remove all leading zeros and the highest bit set of our input value.
The result is then shifted right by 9 bits so that the mantissa occupies the lowest 23 bits.
Placing the 8-bit exponent in the bits 24 to 31 (counting up from the least significant bit) by shifting it left by 23 bits finalizes the mapping.
The sign bit in the position of the most significant bit is always zero.
Comparing the difference between the mapped values and a reference computed with 64-bit precision for all possible 32-bit floating point values confirms that the implementation is indeed optimal for 32 bits.

Note that for all floating point numbers $\geq \frac{1}{256}$
there will be collisions, i.e. multiple integers mapping to the
same floating point number. There is no way around
that due to the logarithmic spacing of the floating point
numbers.

\begin{algorithm}
\begin{lstlisting}
float map_u32_to_unifloat(uint32_t u)
{
    if (u == 0)
        return 0.f; // snap to 0
    if (u == 1u)
        return __uint_as_float((126-31)<<23);
    uint32_t z = __clz(u); // count leading zeros
        // z is <= 30, as u=0 and u=1 already checked
    uint32_t e = 126-z;    // biased exponent
    uint32_t m = (u << (z+1)) >> 9; // discard leading
        // zeroes and first set bit (if any) and
        // shift significant bits into place
    return __uint_as_float((e << 23) | m); // assemble fp32
}

float radinv(uint32_t i, const uint32_t prime_idx)
    // allows for full index range [0..0xFFFFFFFF]
{
    const uint32_t base = primes[prime_idx];
    i %= max_powers[prime_idx]; // limit domain
    uint32_t base_tmp = 1;
    uint32_t result = 0;
 
    do {
        result = result*base + i % base;
        i /= base;
        base_tmp *= base;
    } while(i);
 
    return map_u32_to_unifloat((uint32_t)
        (((uint64_t) result<<32) / base_tmp));
}
\end{lstlisting}
\caption{Routine to optimally map integers in $[0, 2^{32})$ to the
unit interval $[0,1)$ and basic radical inversion. The table \texttt{primes}
stores the prime numbers and \texttt{max\_powers} stores the
largest prime power that still fits into 32 bits.}
\label{Fig:CodeRadicalInversion}
\end{algorithm}

\subsection{Efficient Radical Inversion} \label{Sec:RadInv}

Radical inversion is at the core of most popular low discrepancy
sequences such as the family of Halton, digital $(t,s)$-, and rank-1 lattice sequences.
A radical inverse is computed by the mapping
\begin{eqnarray*}
  \mathbb{N}_0 & \rightarrow & [0,1) \cap \mathbb{Q}\\
    i = \sum_{k = 0}^\infty a_k(i) b^k & \mapsto & \phi_{b, \sigma}(i) = \sum_{k = 0}^\infty \sigma_b(a_k(i))  b^{- k - 1} \, ,
\end{eqnarray*}
where an index $i$ is represented by its digits $a_k$ in base $b$ and
then these digits are mirrored at the decimal point. The resulting sequence of
rational numbers in the right-open unit interval is of low discrepancy~\cite{Nie:92}.
$\sigma_b$ is a permutation on the range of a digit. If $\sigma_b$ is the identity, we will use the shorthand $\phi_b$.

Algorithm~\ref{Fig:CodeRadicalInversion} computes in integers
as long as possible and only converts into floating point as a last step.
This rules out any cancellation effects due to floating point summation.
For efficiency, the index $i$ is limited to the largest prime power
that still fits into 32 bits. These values are stored in the table \texttt{max\_powers},
one for each prime base.

Especially the uniformity of low-dimensional projections of low
discrepancy sequences based on radical inverses can be improved by
applying permutations $\sigma_b$~\cite{Faure:92,LinearDigitScrambledHalton}
(also known as "digit scrambling") during radical inversion.
The routine \texttt{radinv\_linscramble} in Algorithm~\ref{Fig:CodeLinearScramble}
exemplifies computational scrambling by a linear transformation~\cite{LinearDigitScrambledHalton}.
Note how the implementation prevents overflows of integer multiplications.

Small lookup tables can serve both the purpose of storing arbitrary
permutations $\sigma_b$ and more efficient simultaneous inversion of multiple digits~\cite{MCQMCSigCourse2012}
(see Algorithm~\ref{Fig:CodeLinearScramble} for the example of $\phi_3$).
This is illustrated for the recursively defined permutations introduced by Faure~\cite{Faure:92}:
For even $b$, $\sigma_b$ is $2 \sigma_{\frac{b}{2}}$ concatenated with $2 \sigma_{\frac{b}{2}} + 1$.
Otherwise, $\sigma_b$ is $\sigma_{b-1}$, where each value $\geq \frac{b - 1}{2}$ is incremented by 1
and $\frac{b - 1}{2}$ is inserted in the middle. The recursion ends at $\sigma_2 := (0, 1)$. For example, $\sigma_5 = (0,3,2,1,4)$.
In order to compute the radical inversion for two digits, the isomorphism
\[
\sigma_5 \times \sigma_5 =
\begin{pmatrix}
(0,0)&(0,3)&(0,2)&(0,1)&(0,4)\\
(3,0)&(3,3)&(3,2)&(3,1)&(3,4)\\
(2,0)&(2,3)&(2,2)&(2,1)&(2,4)\\
(1,0)&(1,3)&(1,2)&(1,1)&(1,4)\\
(4,0)&(4,3)&(4,2)&(4,1)&(4,4)\\
\end{pmatrix}
\cong
\begin{pmatrix}
0&3&2&1&4\\
25&28&27&26&29\\
10&13&12&11&14\\
5&8&7&6&9\\
20&23&22&21&24\\
\end{pmatrix}
\]
is used and tabulated\footnote{Code for the generation of such tables is found at \url{https://gruenschloss.org}.}. The loop for radical inversion then
performs computations in base $b^2$, requiring only half the number
of iterations. It is straightforward to extend the principle to more
digits. As the size of the lookup table grows exponentially in the
number of digits, this approach pays off only for a moderate
number of digits and is not efficient once the lookup
tables do not stay in the processor caches. Unless the total number of digits to be inverted
is a multiple of the digits inverted simultaneously, the remaining digits
will have to be inverted individually.

Loop unrolling is an extreme optimization yielding branchless
code, which becomes possible by the intrinsically limited number of iterations
of radical inversion. The efficiency of loop unrolling varies according to code size and application
as it may cool the instruction cache. Unless already done by the compiler,
costly division and modulo operations may be replaced by cheaper multiplications, shifts, additions, and subtractions~\cite{HackersDelight}.

\begin{algorithm}
\begin{lstlisting}
float radinv_linscramble(uint32_t i, const uint32_t prime_idx)
{
    const uint32_t base = primes[prime_idx];
    const uint32_t scramble = permTN2[prime_idx];
    i %= max_powers[prime_idx];
    uint32_t base_tmp = 1;
    uint32_t result = 0;
    uint32_t im = i % base; // to avoid scramble*im overflow
 
    do {
        result = result*base + (scramble*im) % base;
        // im % base not required, as im < 0xFFFFFFFF/scramble
        // because base is always > scramble by definition
        i /= base;
        im = i;
        base_tmp *= base;
    } while(i);
 
    return map_u32_to_unifloat((uint32_t)
        (((uint64_t) result<<32) / base_tmp));
}

float radinv3(uint32_t i)
{
    const unsigned char radinv3_perm[] = {0,3,6,1,4,7,2,5,8};
    i %= 3486784401u; // max_powers
    uint32_t base_tmp = 1;
    uint32_t result = 0;
 
    do {          // two digits at once, hence base 9 = 3 x 3
        result = result*9 + radinv3_perm[i % 9];
        i /= 9;
        base_tmp *= 9;
    } while(i);
 
    return map_u32_to_unifloat((uint32_t)
        (((uint64_t) result<<32) / base_tmp));
}
\end{lstlisting}
\caption{Examples of linear digit scrambling and simultaneous multiple-digit radical
inversion.}
\label{Fig:CodeLinearScramble}
\end{algorithm}

\begin{algorithm}
\begin{lstlisting}
float sobol(uint2 i, const uint32_t j, uint32_t scramble = 0)
{
    const uint32_t adr = j*(52/4);
 
    for (uint32_t c = 0; (i.x != 0); i.x>>=4, ++c) {
          const uint4 matrix = sobol_table[adr + c];
          scramble ^= ((matrix.x&((i.x & 1) ? 0xFFFFFFFFu : 0))
              ^(matrix.y&((i.x & 2) ? 0xFFFFFFFFu : 0)))
              ^((matrix.z&((i.x & 4) ? 0xFFFFFFFFu : 0))
              ^(matrix.w&((i.x & 8) ? 0xFFFFFFFFu : 0)));
    }
 
    i.y &= 0xFFFFFu; // only first 20 bits can be looked up,
                     // overall 52bits=32+20
 
    for (uint32_t c = 8; (i.y != 0); i.y>>=4, ++c) {
          const uint4 matrix = sobol_table[adr + c];
          scramble ^= ((matrix.x&((i.y & 1) ? 0xFFFFFFFFu : 0))
              ^(matrix.y&((i.y & 2) ? 0xFFFFFFFFu : 0)))
              ^((matrix.z&((i.y & 4) ? 0xFFFFFFFFu : 0))
              ^(matrix.w&((i.y & 8) ? 0xFFFFFFFFu : 0)));
    }
 
    return map_u32_to_unifloat(scramble);
}
\end{lstlisting}
\caption{CUDA C++ implementation of a selected component of the $i$-th point of a Sobol' sequence.
The generator matrices are stored in \texttt{sobol\_table} with 23 bits of precision, while 64 bits
are used for the index $i$ to enable large numbers of samples. \texttt{0xFFFFFFFFu} means that all 64 bits are set.}
\label{Fig:CodeSobol}
\end{algorithm}

\subsection{Digital Nets and Sequences}

The algorithm
\begin{equation} \label{Eqn:ts}
x_i^{(j)} = \left(\begin{array}{c}b^{-1} \\ \vdots \\ b^{-m}\end{array}\right)^T
\cdot \underbrace{C_j \cdot \left(\begin{array}{c}a_0(i) \\ \vdots \\ a_{m-1}(i) \end{array}\right)}_\text{multiplication in $\mathbb{F}_b$}
  \in [0,1)
\end{equation}
for digital nets and sequences~\cite{Nie:92} of low discrepancy has
the advantage that once implemented, its generator matrices $C_j$
may be replaced by optimized variants (see Sec.~\ref{Sec:Optimization}).

Algorithm~\ref{Fig:CodeSobol} implements the Sobol' sequence~\cite{SobolJK,JK03,JK08},
which is the most popular $(t,s)$-sequence in base $b = 2$. The
$j$-th coordinate $x_i^{(j)}$ is computed by loading 4
columns of the generator matrix $C_j$ at once to perfectly match a memory
interface that fetches 128 bits simultaneously. Each row is masked according
to the corresponding bit of the index $i$ and the result is the bit-wise
\texttt{xor} (vector arithmetic in $\mathbb{F}_2$) of all included columns, computing 4 digits simultaneously.
The generator matrices are stored for 23 bits of floating point
precision, but allow for up to $m = 64$ bit indices in order to enable large numbers
of samples. The bits beyond the least significant 32 bits of the index $i$
are determined in the second part of the code example. The integer result
then needs to be mapped to floating point as described in Sec.~\ref{Sec:fp}.
Since the result is solely computed by \texttt{xor} operations, digit scrambling
can be trivially realized by initializing the parameter \texttt{scramble} with a
random integer rather than with zero.

\subsection{Rank-1 Lattice Sequences}

Given a radical inverse $\phi_b(i)$ and a generator vector $\vec g = (g_0, \ldots, g_{s-1})$,
whose components $g_j =_b \cdots g_{j, 3} g_{j, 2} g_{j, 1} g_{j, 0}$ are infinite
sequences of digits $g_{j,k} \in \{0, \ldots b - 1\}$, a rank-1 lattice sequence \cite{MaizePhD,eLattices:01,Maize:12} consists
of the points
\begin{equation} \label{Eqn:R1LS}
   \vec {x}_i = \left( \phi_b(i) \cdot \vec g \right) \bmod [0,1)^s
\end{equation}
that will be uniformly distributed if the $g_j$ are linearly independent over the rational integers~\cite[Thm. 3.1.3, p. 81]{MaizePhD}.

As in practical applications the number of samples ultimately is finite,
the required least significant digits of the $g_j$ may be represented by integers, yielding a
fast and simple algorithm for generating deterministic uniformly distributed
points.

\subsubsection{Admissible Generator Vectors in Base $b = 2$} \label{Sec:Admissible}

We recollect
properties of admissible generator vectors (also see \cite[Sec.2.1]{PolyR1LS}):
Selecting the components $g_j$ co-prime to the basis $b$,
each dimension is perfectly stratified whenever the number of points is a
power of $b$. The corresponding lattice then is a Latin hypercube sample,
similar to the construction of the Sobol' sequence. In Algorithm~\ref{Alg:R1LS},
where $b = 2$, Latin hypercube samples result from odd $g_j$ for $2^m$ points.

If the components of the generator vector are not unique,
there will exist dimensions that are sampled identically
instead of uniformly. However, uniqueness is not sufficient,
because for $b^m$ points, the components may be identical
modulo $b^m$.
The issue is intrinsic to the construction of rank-1 lattice sequences
and in \cite{PolyR1LS} has been addressed by requiring the components
$g_j \bmod b^{m}$ to be unique for $0 \leq j < b^m$ and any $m \in \mathbb{N}$.
This underlines why the components $g_j$ need to be $b$-adic integers, as otherwise
there exists a number $b^m$ of points from which on
components coincide. In practice, where the $g_j$ are represented
by integers and the number of samples remains finite, these
integers must hence be chosen as large as possible.

Even if the previous constraint is fulfilled, components may be
identical modulo $b^m$, because $g_j \bmod b^m$ generates the same
coordinates as $b^m - (g_j \bmod b^m)$. Similar to before, avoiding this symmetry
is possible for the first $b^{m-1}$ components $g_j \bmod b^m$ and $m \in \mathbb{N}$.
However, once fulfilled for the first $b^{m-1}$ components, it cannot be
fulfilled modulo $b^{m'}$ for any $m' < m$. Avoiding the issue
completely requires generator components of the form $g_j = 2 \cdot 2^j + 1$,
which in turn means that the $g_j$ and the number of points
must grow exponentially with dimension.

And yet there remains a flaw: As only the first some least significant
digits of the lower dimensions are inspected, for large numbers of points,
the more significant digits of the first some generator vector components
have never been inspected: While the  construction of a uniform
rank-1 lattice sequence by primitive polynomials~\cite{PolyR1LS}
in base $b = 2$ yields unique generator vector components,
uniformity still may not be optimal for certain combinations of
number of points and dimensions.

One reason for temporarily compromised uniformity are numbers
$g_j \bmod b^{m}$ that are small as compared to $b^m$. This
can be seen by rewriting the rank-1 lattice sequence in \eqref{Eqn:R1LS}
as sequence of shifted rank-1 lattices $\vec x_{k b^m}, \dots, \vec x_{(k + 1) b^m - 1}, k \in \mathbb{N}_0$,
where
\[
    \phi_b(i + k b^m) \cdot \vec g
      = \left(\phi_b(i) + \phi_b(k b^m)\right) \cdot \vec g
      = \phi_b(i) \cdot \vec g + \underbrace{\phi_b(k) b^{-m-1} \vec g}_{=: \Delta_k}
\]
for $0 \leq i < b^m$.
The shifts $\Delta_k$, who themselves are a rank-1 lattice sequence, are
tiny when the components of $\vec g \bmod b^m$ are small as compared to $b^m$.
As a consequence, the shifted lattices only slowly will fill up the unit cube
uniformly.

There is a second interesting observation that follows from \cite{PolyR1LS}:
Constraining the $g_j \bmod 2^m$ to be odd and unique for
$0 \leq j < 2^{m-1}$ and $m \in \mathbb{N}$, obviously any first $2^{m-1}$
components $g_j \bmod 2^m$ are a permutation of all odd natural numbers
less than $2^m$. Simply selecting $g_j = 2 j + 1$ does not result in
good lattices, when the $g_j$ are small with respect to $b^m$ as explained above. 
In other words, finding a good generator vector is equivalent
to reordering (or enumerating) the admissible components as mentioned in \cite{PolyR1LS} and
investigated in graphics~\cite{R1Lpathintegral}.

\begin{algorithm}
\begin{lstlisting}
float random_lattice(uint32_t i, uint32_t j,
    uint32_t pixel_pos_x, uint32_t pixel_pos_y)
{
    uint32_t hash
        = hash_value(j, pixel_pos_x, pixel_pos_y);
    uint32_t lattice_param = hash | 1;
   
    uint32_t result = bit_reversal(0xFFFFFFFFu-i) // __brev
        * lattice_param; // backwards, to avoid 0
    return map_u32_to_unifloat(result);
}\end{lstlisting}
\caption{Given pixel coordinates, \texttt{random\_lattice} computes
the selected component $j$ of the $i$-th point of a rank-1 lattice sequence.
The generator vector component is computed as a pseudo-random hash of the pixel position
and component index.
Note that the points are enumerated backwards to
avoid sampling the common origin first. Ingnoring the 32 most signficant
bits in the integer multiplication amounts to the $\bmod \, 1$ operation.}
\label{Alg:R1LS}
\end{algorithm}

\subsubsection{Admissible Random Generator Vectors in Base $b=2$}

While good generator vectors do exist \cite{Hick:02}, there are no constructions
in high dimensions and an exhaustive search is infeasible. For certain
smoothness classes of functions, the component-by-component construction
\cite{LatSeq06} has become practical.
Interestingly, selecting generator vectors at random~\cite{RandomLattices,RandomMedianLattices} has shown
quite good results. Such construction-free approaches avoid a computer
search and lend themselves to problems that do not provide structure other than
square integrability, like for example, light transport simulation.

As an example from computer graphics, Algorithm~\ref{Alg:R1LS} determines
the coordinate $x_i^{(j)}$ according to \eqref{Eqn:R1LS} by computing a hash value
from the dimension $j$ and the pixel coordinates in an image. Setting the
least significant bit of the pseudo-random value yields the generator vector
component $g_j$. Forcing the components to be odd results in perfect
stratification along each dimension for a base $b = 2$.
This way, each pixel is assigned its own, pseudo-random
rank-1 lattice sequence generator vector. Furthermore, the samples are
enumerated backwards in order to avoid sampling with the origin as a
common first sample across the screen, as this would result in aliasing
rather than noise. Yet, the zero is not excluded~\cite{OwenFirstPoint} -- it only comes last.
The implementation is deterministic and hence reproducible. It does not
access any lookup tables, which avoids any latency due to memory access.

As an alternative, the sequence of $g_j = 2 \xi +1$ may be generated by a pseudo-random
number generator $\xi$, for example, a linear feedback shift register (LFSR) generator.
The seed of the generator then defines the generator vector.
Since $\xi = 0$ is not element of the LFSR sequence, a component with $g_j = 1$
can be added, which amounts to the radical inverse $\phi_2$ in that dimension.
As compared to \cite{PolyR1LS}, the algorithm uses only one LFSR instead
of one per dimension and loads only one seed that determines the generator
vector components computed on the fly without any further memory access.
It is hence a highly desirable implementation on modern processors.

Creating an admissible random generator vector with unique components
by using a pseudo-random number generator is simple. Guaranteeing
uniqueness for the hash-based approach is tricky and in the extreme
case requires computational verification. Another caveat is that it is
not obvious how to enforce the criteria from the previous section when
creating pseudo-random generator vectors. In a straightforward approach,
one would skip pseudo-random numbers not fulfilling the criteria,
which, for efficiency reasons, would require to tabulate the generator
vectors.

\subsection{Massive Parallelization} \label{Sec:Parallel}

There are two key aspects of parallelization: Solving a large problem
by parallel computation and solving a large number of problems in parallel.
Quasi-Monte Carlo methods cover both aspects, are efficient in massively parallel
heterogenous computing environments, and in addition do not require intermediate
synchronization.

Computing several millions of integrals simultaneously, 32-bit indices may not
be sufficient. In addition, errors caused by an index wrapping around in 32 bits
may be subtle and hard to find. It hence is important to be certain about
the range of an index required by an application.

\subsubsection{Partitioning Low Discrepancy Sequences}

One approach to parallelization is to partition one low discrepancy
sequence into multiple low discrepancy sequences and to assign each
one subsequence to a processing element~\cite{ParQMC}. To do so, a
low discrepancy sequence is extended by one dimension. The unit
interval corresponding to the additional dimension is partitioned into
the number $P$ of processing elements. All points of the low discrepancy
sequence whose additional dimension is element of the $p$-th
interval are assigned to the $p$-th processing element. For low
discrepancy sequences based on radical inversion, the subsequences
can be efficiently enumerated per processing element~\cite{ParQMC}.

A second approach to massive parallelization has been developed for
integro-approximation in computer graphics~\cite{SampleEnum,Iray_report}. In order to compute
the color of each pixel in an image, a high-dimensional integral needs
to be computed~\cite{PBRT}. Using the first two dimensions of one
high-dimensional low discrepancy sequence, the stratification imposed
by the pixels in the image plane is used to assign each point of the
low discrepancy sequence to exactly one pixel. Like before, the
subsequence corresponding to one pixel can be efficiently
enumerated for radical inversion based low discrepancy sequences.

Parallelization by partitioning low discrepancy sequences begins
by assigning the subsequences to processing elements and ends
with the reduction of the results. No intermediate communication
is required. Since each processing element can enumerate its
own subsequence, progressive and adaptive sampling can be
controlled independently of the other processing elements.

Each part of the computation is deterministic and hence reproducible
in principle. However, it is not always possible to control the exact order
of additions in a massively parallel computing environment. This may
introduce cancellation effects of random nature when repeating
computations. To control this source of randomness, cancellation
effects need to be kept at a minimum, for example by Kahan's
algorithm, hierarchical reduction, sorting numbers before addition, or computation in integers, the last of which is truly commutative and associative as long as the results do not leave the integer range.
A second caveat of the partitioning approach is that the number
of required points in a massively parallel computation may easily
exceed an index range representable within 32 bits. This can be
resolved by using 64-bit integers for the indices, which comes at
a certain cost, especially when 64-bit operations are emulated by 32 bit operations.

\subsubsection{Randomized Quasi-Monte Carlo Methods}

By randomizing one low discrepancy sequence~\cite{Owen:98,lecuyer02}, each processing element
may generate an individual low discrepancy sequence. This keeps the
index range reasonable, even with massive parallelization. The reduction
of the single results allows for an unbiased variance estimate, since
all randomized sequences are independent. This comes at the price
that the union of samples very unlikely is of low discrepancy, falling
short with respect to maximum efficiency.

While this may be an issue when massively parallelizing one big
problem, derandomized randomized quasi-Monte Carlo methods
are an opportunity when computing many integrals in
parallel as addressed in the next section on optimization.

\subsection{Optimization} \label{Sec:Optimization}

The algorithms for low discrepancy sequences based on radical inversion,
generator matrices, and generator vectors share the property that once an efficient
implementation is in place, the permutations and generators may be optimized
and adapted to the actual application.

It has been recognized early on that random or randomized generator
matrices \cite{ReginaPhD} perform quite well and often better than the classic
deterministic constructions~\cite{Myths04}. Hence, instead of searching for better
number-theoretic constructions, we may take advantage of the fact that many
sets of random parameters are good and single out the bad ones. One such approach
is to use the median of a number of randomized quasi-Monte Carlo
integrations~\cite{UniversalMedianQMC} that inspired investigations using
random generator matrices~\cite{RandomGeneratorMatrices} and random
generator vectors~\cite{RandomMedianLattices}.

In certain applications,  the efficiency of using the median of repeated
integration may be disputable and for this reason finding parameters
by computer search remains an option. Minimizing the overall
discrepancy and thus improving uniformity does not necessarily
result in good low dimensional projections. This has been addressed
in~\cite{JK08} for the Sobol' sequence and led to a more general
framework to specify $(t,m,s)$-nets and $(t,s)$-sequences by constraints
on projections~\cite{MatBuilder}. This way, uniform point sets can be
tailored to match lower-dimensional structures intrinsic to
an integration problem. This is of advantage in finance and physics,
where high-dimensional integrals often are composed of a sequence
of low dimensional integrals. Similar tools exist for lattice rules~\cite{l2022tool}.

Maximizing the minimum distance of point sets~\cite{Myths04} is
meaningful in small dimensions, especially in graphics. Corresponding
experiments~\cite{PermNets} even reveal the existence of $(t,m,s)$-nets that
cannot be represented by \eqref{Eqn:ts}. A related approach allows
for designing a point set according to a desired energy spectrum~\cite{DeepPointCorrelationDesign}.
While these approaches have benefits in two dimensions, they do not
translate to efficient high-dimensional approaches.

Using higher bases and prime power bases in radical inversion (see Sec.~\ref{Sec:RadInv})
provides a larger design space for permutations and generator
matrices. This is especially interesting to explore with computational
scrambling.

When no means are at hand
to assess the discrepancy of a point sequence, its quality may be inspected by
integrating test functions. Many practical test functions have analytic integrals
and are designed to be sensitive to irregularities of distribution when integrated
numerically~(for example, see \cite[Sec.5.1]{WeightedCompoundIntegration}).
Sometimes test functions are invariant to permutations of the dimensions, which,
however, can make a difference in applications.
As elaborated in Sec.~\ref{Sec:Admissible} and explored in~\cite{R1Lpathintegral},
assigning the dimensions of a low discrepancy sequence to the problem
dimensions matters. This amounts to assigning the best uniformly distributed
dimensions of a low discrepancy sequence to the most important dimensions
of a problem. In practice, problems like
light transport simulation typically act like a lens on the quality of low dimensional
projections of a point sequence, as they reveal a lack of uniformity
by exposing visual artifact structure in images.

\subsubsection{Two Optimization Examples from Computer Graphics}

Real-time graphics allows for only a very few samples per pixel to estimate
its color. The resulting images likely look noisy and hence need to be filtered.
Interestingly, the characteristic of the noise makes a difference. In fact, blue
noise is most amenable to the human eye, as frequencies below the Shannon
limit can be reproduced and frequencies above are mapped to noise, reducing
distracting visible aliasing.
This has been known for long, especially in computer graphics, however,
only recently the focus shifted from blue noise properties
of samples within a pixel to blue noise across the image plane~\cite{DitheredSampling}.

Subsequent work has led to a very efficient implementation~\cite{ScreenSpaceBlueNoise}
that uses one low discrepancy sequence across all pixels and per-pixel randomization
to achieve blue noise characteristics rather than aliasing:
The tiny algorithm stores $s$-dimensional points $P = (p_1, \ldots, p_N)$ of an Owen-scrambled
Sobol' sequence (see Algorithm~\ref{Fig:CodeSobol}) as integers along with a digit scrambling table $s_{x,y}$ of
$128^2 \times s$ integers and a scrambling table $r_{x,y}$ of $128^2$
integers for reordering the points. Given the pixel coordinates $(x,y) \bmod 128$,
the $i$-th sample is determined by
\[
  p_i^{(x,y)} = p_{i \oplus r_{x,y}} \oplus s_{x,y} \, ,
\]
where $\oplus$ is the bit-wise exclusive-or operation. The resulting integer $p_i^{(x,y)}$ is converted to floating
point (see Algorithm~\ref{Fig:CodeRadicalInversion}). Both scrambling tables
have been found by minimizing an energy functional by simulated annealing
that yields the blue noise characteristics across the pixels.

Following these principles and inspecting the structure of light transport simulation by path tracing,
an even simpler sampler~\cite{ScreenSpaceBlueNoise2} has been found
shortly after. Tracing a light transport path consists of simulating a sequence of
scattering events and visibility tests. The important observation is that
aliasing will appear if at a certain bounce samples are correlated across pixels
rather than along the whole path. Hence, to achieve blue noise properties,
only one $128^2$ table of two-dimensional shift vectors for Cranley-Patterson-rotations
that is repeated for all pairs of dimensions is sufficient to decorrelate
high-dimensional samples of one rank-1 lattice sequence with generator
vectors from~\cite{WeightedCompoundIntegration} across the pixels.
The table of shifts has been computed by minimizing the variance across a set of test functions,
where the loss has been determined after denoising the test images. Hence
the shifts are already optimized with respect to noise filtering in mind.

As compared to the first example~\cite{ScreenSpaceBlueNoise}, the points of the rank-1 lattice sequence can be
generated without a lookup table and the remaining lookup table for the
Cranley-Patterson-rotations is small and low-dimensional, rendering its
optimization much less complex~\cite{ScreenSpaceBlueNoise2}. Note that the design of the sampling
algorithm is closely coupled to the structure of image synthesis by
light transport simulation, i.e. the structure of the Neumann series
to solve an integral equation.

\begin{figure}
\noindent{\renewcommand{\arraystretch}{0.0}%
\setlength{\tabcolsep}{0pt}%
\setlength{\arrayrulewidth}{0pt}%
\def\colw{.2375\textwidth}%
\begin{tabular}{@{}p{.05\textwidth}c@{}}%
\tikz{\node[minimum width=705/1920*.95*\textwidth,inner sep = 0, outer sep = 0, rotate=90] at (0, 0) { Reference };}&%
\begin{tikzpicture}
\node[anchor=north west, inner sep = 0, outer sep = 0] at (0, 0) {\includegraphics[width=.95\textwidth]{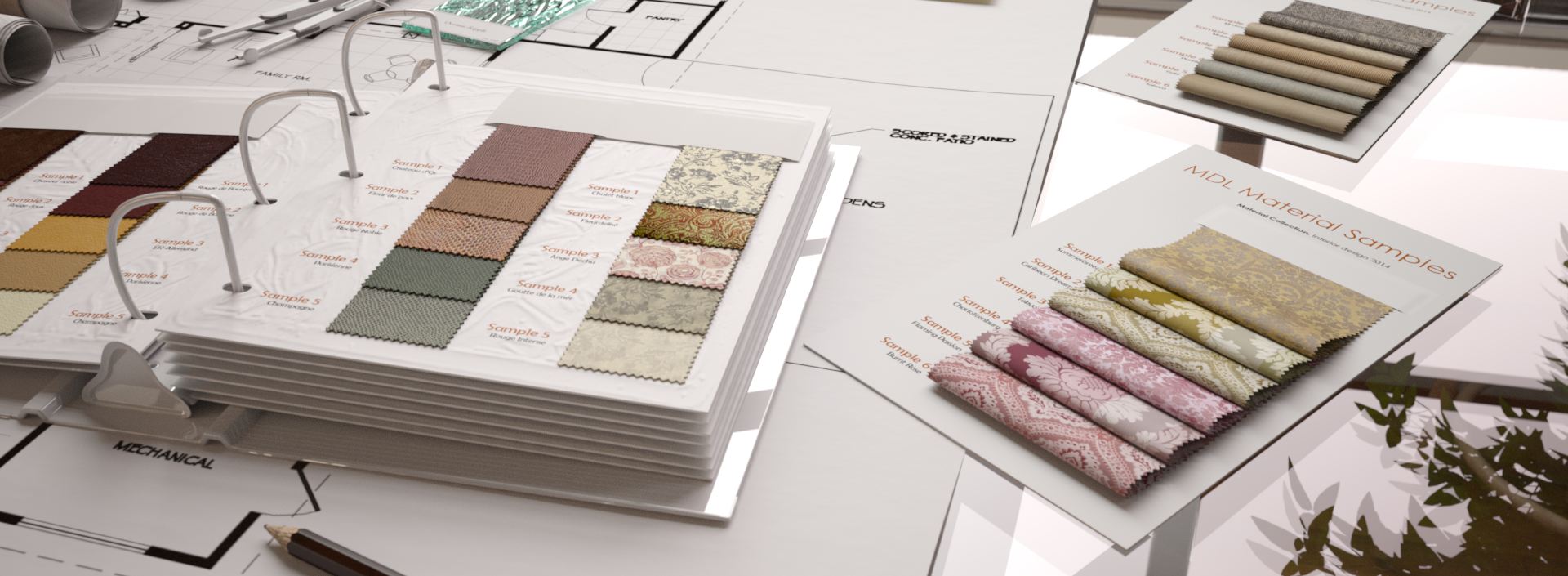}};
\draw[HighlightColor, very thick] (379/1920*0.95*\textwidth, -837/1920*705/1080*0.95*\textwidth) rectangle ++(252/1920*0.95*\textwidth, -129/1920*0.95*\textwidth);
\end{tikzpicture}%
\end{tabular}\\%
\begin{tabular}{@{}p{.05\textwidth}cccc@{}}
\tikz{\node[minimum width=129/252*\colw,inner sep = 0, outer sep = 0, rotate=90] at (0, 0) { 256 spp };}&%
\includegraphics[width=\colw]{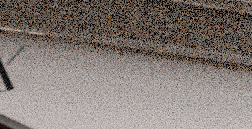}&%
\includegraphics[width=\colw]{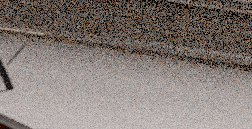}&%
\includegraphics[width=\colw]{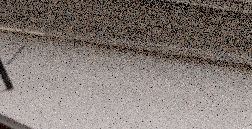}&%
\includegraphics[width=\colw]{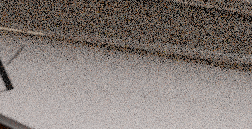}\\
\tikz{\node[minimum width=129/252*\colw,inner sep = 0, outer sep = 0, rotate=90] at (0, 0) { 64 spp };}&%
\includegraphics[width=\colw]{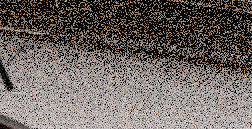}&%
\includegraphics[width=\colw]{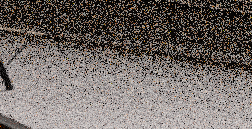}&%
\includegraphics[width=\colw]{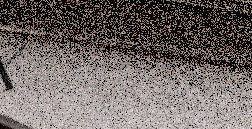}&%
\includegraphics[width=\colw]{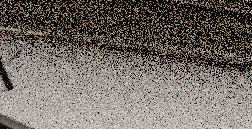}\\
\tikz{\node[minimum width=129/252*\colw,inner sep = 0, outer sep = 0, rotate=90] at (0, 0) { 16 spp };}&%
\includegraphics[width=\colw]{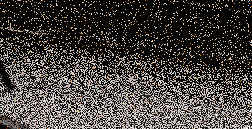}&%
\includegraphics[width=\colw]{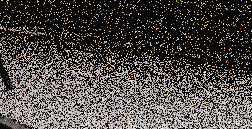}&%
\includegraphics[width=\colw]{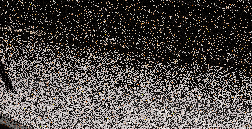}&%
\includegraphics[width=\colw]{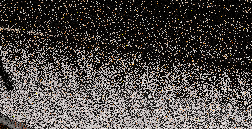}\\
\tikz{\node[minimum width=129/252*\colw,inner sep = 0, outer sep = 0, rotate=90] at (0, 0) { 4 spp };}&%
\includegraphics[width=\colw]{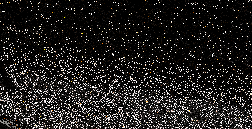}&%
\includegraphics[width=\colw]{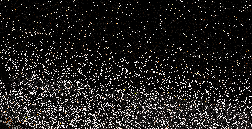}&%
\includegraphics[width=\colw]{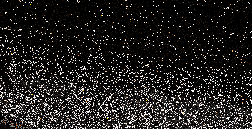}&%
\includegraphics[width=\colw]{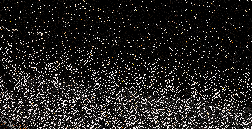}\\[.5em]
& Halton along Hilbert &
Pixel shifted lattice &
Pixel random lattice &
Image plane Halton \\[.5em]
& as in \cite{AlongHilbert}
& as in Eq.~\eqref{Eqn:PixelShiftedR1LS}
& as in Algorithm~\ref{Alg:R1LS}
& as in \cite{SampleEnum}
\end{tabular}}
\caption{Evolution of noise characteristics in light transport simulation at low number of samples per pixel (spp).
The pixel shifted lattice as in Eq.~\eqref{Eqn:PixelShiftedR1LS} shows the overall best performance. At 256 spp, its noise is
most uniform, has the least amount of black pixels, and does not expose structural correlation artifacts. Lacking
guarantees on the quality of the generator vector, the pixel random lattices still show black pixels, exposing
bad generator vectors. At low sampling rates, especially visible at 16 spp and 4 spp, the approaches using the
Halton sequence expose axis-aligned structures, streaks, and ripples.}
\label{Fig:Results}
\end{figure}

\section{Results} \label{Sec:Results}

With the teachings of the previous two examples from
computer graphics,  the question arises whether a similar
reduction in the perceptual error can be achieved without optimization.
In~\cite{AlongHilbert} multiple sampling algorithms have been
introduced that take advantage of enumerating low discrepancy
sequences along the Hilbert curve. In fact, they achieve
characteristics similar to those of the desirable blue noise
characteristics~\cite{DitheredSampling,ScreenSpaceBlueNoise,ScreenSpaceBlueNoise2}
without optimization.

In the following, we extend these investigations to new and simple
algorithms based on rank-1 lattice sequences in base $b= 2$ as they only require
three operations: bit reversal to compute $\phi_2$, multiplication,
and mapping to floating point numbers in the range $[0, 1)$ as
shown in Algorithm~\ref{Fig:CodeRadicalInversion}.
Fig.~\ref{Fig:Results} compares the results of Algorithm~\ref{Alg:R1LS} to
prior work of enumerating a Halton sequence along the Hilbert curve
over the image plane~\cite{AlongHilbert} and sampling the image plane
by using the first two dimensions of a Halton sequence~\cite{SampleEnum}.
While the latter algorithms are consistent and hence convergent,
at low sampling rates, the noise characteristics of the rank-1 lattice
sequence algorithm outperform the image plane sampling approach.
This is remarkable, as the extremely simple Algorithm~\ref{Alg:R1LS}
uses a pseudo-random generator vector per pixel and is more efficient
to generate than the Halton sequence.

The outliers at 256 samples per pixel that become visible as salt-and-pepper
noise are due to bad instances of the pseudo-random generator vectors. We
find their number so small that a denoiser can remove them. As an alternative,
already superimposing four instances of the random rank-1 lattice sequence
in a pixel reliably hides the outliers, while remaining competitive in terms
of performance. This favorably relates to removing outliers by a median
\cite{UniversalMedianQMC,RandomGeneratorMatrices,RandomMedianLattices}.

Using one generator vector for all pixels, Belcour and Heitz~\cite{ScreenSpaceBlueNoise2}
uncorrelated pixels by an optimized, padded low dimensional Cranley-Patterson-rotation
table (see the previous section). In this paper, we propose the following, similar but new
algorithm which does not require any optimization:
\begin{equation} \label{Eqn:PixelShiftedR1LS}
  \left(\phi_2\left(i\right) + \phi_3\left(\texttt{inv\_hilbert}\left( x, y \right)\right)\right) \cdot g_j\ \textrm{mod}\ 1
\end{equation}
Instead of applying a single shift to each dimension of the rank-1 lattice sequence in Eq.~\eqref{Eqn:R1LS}, only one one-dimensional shift
is applied to the radical inverse $\phi_2$ before the generator vector multiplication.
Drawing from prior work~\cite{AlongHilbert}, we determine this one shift per pixel by
enumerating the radical inverse $\phi_3$ along the Hilbert curve in the image plane,
where $(x, y)$ are the pixel coordinates. This preserves the low discrepancy of the shifts
across the pixel neighborhoods~\cite{SQMC}. Selecting a co-prime radical inverse
avoids interference with the radical inverse enumerating the points of the
rank-1 lattice sequence. The implementation remains compact and simple: We tabulate
one generator vector $\vec g$ according to Sec.~\ref{Sec:Admissible} and \cite{PolyR1LS}.
The radical inverse $\phi_2$ is computed by bit reversal (see Algorithm~\ref{Alg:R1LS}), while the second radical
inverse $\phi_3$ is implemented using a small lookup table for the simultaneous inversion of four digits
as shown in Algorithm~\ref{Fig:CodeLinearScramble} and derived in Sec.~\ref{Sec:RadInv}.
The resulting code runs about 7 times faster than our previous approach~\cite{SampleEnum},
where we applied the linearly scrambled Halton sequence as in Algorithm~\ref{Fig:CodeLinearScramble},
used permutation tables for the first 32 dimensions in combination with the simultaneous inversion of multiple digits,
and replaced divisions and modulo operations by cheaper operations~\cite{HackersDelight}.

The pixel shifted lattice results in Fig.~\ref{Fig:Results} compare well
to the state of the art: At low sampling rates, they lack axis-aligned structures, streaks, and ripples
of approaches using the Halton sequence. At higher sampling rates, the noise characteristics
are most uniform and avoid the failure cases inherent with randomly selected generator vectors.
For typical use cases with 200-2000 samples per pixel and
for scenes with contemporary geometric complexity, sampling according to Eqn.~\eqref{Eqn:PixelShiftedR1LS}
has been extremely robust while still exhibiting the desirable noise characteristics across sampling rates
and almost minimal code complexity.

\section{Conclusion}

We present quasi-Monte Carlo algorithms for radical inversion, digital
nets and sequences, and rank-1 lattice sequences as they are implemented
in industry for reasons of efficiency, precision, and robustness. With a focus on
rank-1 lattice sequences, current and future optimizations of low discrepancy
sequences have been discussed. Along these lines, a new improved sampling
algorithm for image synthesis has been introduced.
Although developed in the context
of massively parallel light transport simulation, the techniques
in this paper have applications not only in computer graphics software.
An avenue of future research is crafting low discrepancy sequences according
to the intrinsic structure of integro-approximation problems~\cite{MatBuilder}.

\end{document}